\documentclass[showpacs,amsmath,amssymb,aps,pra]{revtex4}
\usepackage{bm}
\usepackage{amsfonts}
\usepackage{graphicx}
\begin{document}

\title{Beams of electromagnetic radiation carrying angular momentum:\\ The Riemann-Silberstein vector and the classical-quantum correspondence}
\author{Iwo Bialynicki-Birula}
\email{birula@cft.edu.pl}
\affiliation{Center for Theoretical Physics, Polish Academy of Sciences, Al. Lotnik\'ow 32/46, 02-668 Warsaw, Poland and\\
Institute of Theoretical Physics, Warsaw University}
\author{Zofia Bialynicka-Birula}
\affiliation{Institute of Physics, Polish Academy of Sciences,\\ Al. Lotnik\'ow 32/46, 02-668 Warsaw, Poland}

\begin{abstract}
All beams of electromagnetic radiation are made of photons. Therefore, it is important to find a precise relationship between the classical properties of the beam and the quantum characteristics of the photons that make a particular beam. It is shown that this relationship is best expressed in terms of the Riemann-Silberstein vector --- a complex combination of the electric and magnetic field vectors --- that plays the role of the photon wave function. The Whittaker representation of this vector in terms of a single complex function satisfying the wave equation greatly simplifies the analysis. Bessel beams, exact Laguerre-Gauss beams, and other related beams of electromagnetic radiation can be described in a unified fashion. The appropriate photon quantum numbers for these beams are identified. Special emphasis is put on the angular momentum of a single photon and its connection with the angular momentum of the beam.
\end{abstract}
\maketitle

\section{Introduction}

{\em ``The notion of photon as a quantum of electromagnetic energy, like the notion of atom as an elementary unit of matter, permits a number of interpretations.''} With these words Bruce Shore begins the section entitled ``What is a photon'' in his comprehensive monograph \cite{bws} and then he goes on to present a few possible answers to this question. We hope that our contribution to this special issue adds a little to the neverending discussion on the nature of photons.

Electromagnetic radiation, especially in the optical range, is produced and used most often in the form of beams. There is a variety of mathematical models to describe such beams: the Bessel beams, the Hermite-Gauss beams, the Laguerre-Gauss beams, and also the focus wave modes. The mathematical representations of these beams are the exact, or approximate, solutions of the classical Maxwell equations. In recent years, there were many experiments that exhibited the influence of the orbital momentum on the beam properties (phase dislocations, optical vortices) and on the interaction of beams with matter (particle trapping, optical tweezers and spanners). As a rule, orbital angular momentum leads to vortices and electromagnetic beams with vortices, as we shown recently \cite{beams,bbc}, are interesting because they can guide charged particles. The fundamental theoretical and experimental papers on the optical angular momentum were recently reprinted in a collection \cite{oam}. In all these papers, the theoretical description of beams carrying angular momentum is given in terms of the classical solutions of the Maxwell equations and sometimes separately in terms of photons. However, no unifying principle is given that would connect these two points of view in a precise manner. The purpose of this work is to fill this gap. We shall show that the notion of the photon wave function appears in both of these descriptions and that it provides a very convenient concept to unify the two points of view. A very useful mathematical tool in this analysis is the Riemann-Silberstein vector \cite{weber,sil,bateman,stratton,qed,app,pio}. Applications of the RS vector to many physical problems were recently reviewed in a very thorough paper by Keller \cite{ok}. The Whittaker representation \cite{whitt} of this vector greatly simplifies the calculations since then the vector field is described by a single function. We shall not consider here the beams described by {\em approximate} solutions of Maxwell equations obtained in the paraxial approximation \cite{apb}. The analysis of such solutions in terms of photons would be rather awkward --- the notion of an approximate photon does not make sense.

There is some overlap between our final conclusions and those obtained recently by J{\'a}regui and Hacyan \cite{jh} even though our methods are completely different. They rely on the standard description of the quantized electromagnetic field based on the vector potential. We follow the methods developed in our earlier works on quantum electrodynamics \cite{qed} and photon wave functions \cite{pio} where the Riemann-Silberstein vector plays the central role.

\section{Succinct description of the electromagnetic field}\label{two}

A natural tool in the analysis of the solutions of the Maxwell equations, in both classical and quantum theories, is the Riemann-Silberstein (RS) vector ${\bm F}$
\begin{eqnarray}\label{rs}
{\bm F} = \sqrt{\frac{\epsilon_0}{2}}({\bm E}+ i c{\bm B}).
\end{eqnarray}
The physical significance of the RS vector has been recognized by Silberstein \cite{sil} who observed that important characteristics of the electromagnetic field (energy density, Poynting vector, Maxwell stress tensor) are bilinear products of the components of this vector. The total energy
\begin{eqnarray}\label{energy}
E = \int\!d^3r\,{\bm F}^*\!\cdot\!{\bm F},
\end{eqnarray}
the total momentum (i.e. the integral of the Poynting vector divided by $c$)
\begin{eqnarray}\label{momentum}
{\bm P} = \frac{-i}{c}\int\!d^3r\,{\bm F}^*\times{\bm F},
\end{eqnarray}
and the total angular momentum
\begin{eqnarray}\label{angmomentum}
{\bm M} = \frac{-i}{c}\int\!d^3r\,\left({\bm r}\times({\bm F}^*\times{\bm F})\right)
\end{eqnarray}
look very much like the quantum-mechanical expectation values and we shall show later that this fact has a deeper meaning.

The convenience of using the RS vector has been recognized by Bateman \cite{bateman}, who was the first to analyze with its help various solutions of the Maxwell equations. Kramers \cite{kramers} used the RS vector to formulate the canonical theory of the electromagnetic field and Power \cite{power} stressed the usefulness of this vector in the description of circularly polarized waves. The complex RS vector carries exactly the same information as two real field vectors but its use significantly simplifies the mathematical analysis. In particular, the two pairs of real Maxwell equations written in terms of ${\bm F}$ reduce to just one pair of complex Maxwell equations
\begin{eqnarray}\label{max}
\partial_t{\bm F}({\bm r},t) = -i c\nabla\times{\bm F}({\bm r},t),\;\;\nabla\!\cdot\!{\bm F}({\bm r},t)=0.
\end{eqnarray}
The RS vector can be expressed in the following form \cite{loc}
\begin{eqnarray}\label{hertz}
{\bm F}({\bm r},t) = \nabla\times\left(\frac{i}{c}\partial_t{\bm Z}({\bm r},t)+\nabla\times{\bm Z}({\bm
r},t)\right),
\end{eqnarray}
where ${\bm Z}({\bm r},t)$ is a complex vector field (a unified form of the two Hertz vector potentials) satisfying the d'Alembert equation
\begin{eqnarray}\label{weq}
(\frac{1}{c^2}\partial_t^2-\Delta){\bm Z}({\bm r},t)=0.
\end{eqnarray}
In the description of beams, it is convenient to choose the vector ${\bm Z}$ in the direction of propagation ${\bm Z}({\bm r},t)=(0,0,1)\chi({\bm r},t))$. In this way, we obtain the following representation of the RS vector in terms of one complex function $\chi({\bm r},t)$
\begin{subequations}\label{solf}
\begin{eqnarray}
F_x&=&(\partial_x\partial_z + \frac{i}{c}\partial_y\partial_t)\chi({\bm r},t),\\
F_y&=&(\partial_y\partial_z - \frac{i}{c}\partial_x\partial_t)\chi({\bm r},t),\\
F_z &=& -(\partial_x^2+\partial_y^2)\chi({\bm r},t).
\end{eqnarray}
\end{subequations}
We shall refer to these formulas as the Whittaker representation. A century ago Whittaker discovered \cite{whitt} that an electromagnetic field obeying the Maxwell equation can be described by two real functions. By separating Eqs.~(\ref{solf}) into the real and imaginary part, we recover the original Whittaker's formulas. The Whittaker representation (\ref{solf}) of the RS vector is the simplest but it is not unique. For example, by choosing the complex Hertz vector ${\bm Z}$ in the form ${\bm Z}({\bm r},t)=(1,i,0)\chi({\bm r},t))$ we obtain a different expression for ${\bm F}$ that will be used in Section \ref{momr}.

The representation of the solutions of the Maxwell equations in terms of a single complex function (or two real functions for that matter) satisfying the d'Alembert equation greatly simplifies the analysis. This has been noticed already by Stratton \cite{stratton} in his derivation of the formulas for the Bessel beams. Since there are no auxiliary conditions (like the vanishing of the divergence) imposed on $\chi$, this single function describes true degrees of freedom of the electromagnetic field. Sometimes, following Whittaker, we shall call $\chi$ a ``scalar'' solution of the d'Alembert equation. However, strictly speaking this terminology is not justified because $\chi$ has fairly complicated transformation properties, quite different from those of a scalar field.

Every solution of the d'Alembert equation can be decomposed into plane waves. Therefore, we can write
\begin{eqnarray}\label{super}
\chi({\bm r},t) = \int\!\!d{\bm k}\,N({\bm k})\left(f^+({\bm k})e^{-i\omega_k t+i{\bm k}\cdot{\bm
r}}\!+f^-({\bm k})e^{i\omega_k t-i{\bm k}\cdot{\bm r}}\right),
\end{eqnarray}
where $\omega_k=k c$, $N({\bm k})$ is a normalization factor, and
\begin{eqnarray}\label{def}
d{\bm k}=\frac{d^3\!k}{(2\pi)^3}.
\end{eqnarray}
With the use of Eqs.~(\ref{solf}), we obtain the following decomposition of the RS vector into plane waves
\begin{eqnarray}\label{super1}
{\bm F}({\bm r},t) = \int\!\!d{\bm k}\,{\bm e}({\bm k})\left(f^+({\bm k})e^{-i\omega_k t+i{\bm k}\cdot{\bm
r}}+f^-({\bm k})e^{i\omega_k t-i{\bm k}\cdot{\bm r}}\right),
\end{eqnarray}
where the complex polarization vector
\begin{eqnarray}\label{polar}
{\bm e}({\bm k}) = N({\bm k})\left(
\begin{array}{c}
-k_xk_z+i kk_y\\
-k_yk_z-i kk_x\\
k_x^2+k_y^2
\end{array}
\right)
\end{eqnarray}
is a normalized solution of the following set of algebraic equations
\begin{eqnarray}\label{poleq}
{\bm n}\times{\bm e}({\bm k})=-i{\bm e}({\bm k}),\;\;\;{\bm n}={\bm k}/k.
\end{eqnarray}
We choose the normalization factor $N({\bm k})$ as
\begin{eqnarray}\label{nrm}
N({\bm k}) = \frac{1}{\sqrt{2}\,kk_\perp},\;\;k_\perp = \sqrt{k_x^2+k_y^2},
\end{eqnarray}
to make ${\bm e}({\bm k})$ normalized to one (${\bm e}({\bm k})\cdot{\bm e}^*({\bm k})=1$). Eqs.~(\ref{poleq}) together with the normalization condition define the polarization vector up to an arbitrary ${\bm k}$-dependent phase. This reflects the gauge freedom of the complex Hertz potential ${\bm Z}({\bm r},t)$.

The two arbitrary complex amplitudes $f^\pm({\bm k})$ describe the true degrees of freedom of the electromagnetic field. They will be labeled with the index $\sigma$ taking on the values $\pm 1$. The sign of $\sigma$ determines whether the pair of vectors $({\bm E},{c\bm B})$ obtained from the plane wave solutions ${\bm e}({\bm k})e^{-i\sigma(\omega t-i{\bm k}\cdot{\bm r})}$ of Maxwell equation (\ref{max}) rotates clockwise ($\sigma$ = 1) or anticlockwise ($\sigma$ = -1) along the direction of propagation. Thus, the amplitudes $f^+({\bm k})$ and $f^-({\bm k})$ describe two circular polarization components of the wave.

\section{Bessel beams}

In the previous section we have given the solutions of the Maxwell equations in the plane wave basis. Since we are interested in the beams with angular momentum, we shall expand the plane waves appearing in (\ref{super}) according to the formula (Eq.~8.511.4 of Ref.\cite{gr})
\begin{eqnarray}\label{expbess0}
e^{i{\bm k}\cdot{\bm r}}=e^{i k_z z}\sum_{m=-\infty}^{\infty}\!i^m e^{i m(\phi-\varphi)} J_m(k_\perp\rho),
\end{eqnarray}
where we introduced a convention that $\phi$ is the polar angle in the $xy$ plane and $\varphi$ is the polar angle in the $k_xk_y$ plane. With the use of this formula, we obtain an expansion of $\chi({\bm r},t)$ into the solutions of the d'Alembert equation in cylindrical coordinates
\begin{eqnarray}\label{expbess}
{\chi}({\bm r},t) = \sum_{m=-\infty}^{\infty}\int_{-\infty}^{\infty}\!\!\frac{d
k_z}{2\pi}\int_{0}^{\infty}\!\frac{d k_\perp k_\perp}{2\pi} \Big(\chi_{k_\perp m k_z}^{+}(\rho,\phi,z,t)
f_{k_\perp m k_z}^+ +\chi_{k_\perp m k_z}^{-}(\rho,\phi,z,t)f_{k_\perp m k_z}^-\Big),
\end{eqnarray}
where
\begin{eqnarray}\label{bes}
\chi_{k_z k_\perp m}^{\sigma}(\rho,\phi,z,t)
 = \frac{(i\sigma)^m}{\sqrt{2}kk_\perp}e^{-i\sigma(\omega_kt- k_z z - m\phi)}J_{m}(k_\perp\rho),
\end{eqnarray}
and
\begin{eqnarray}\label{fcomp}
f_{k_\perp m k_z}^\sigma = \frac{1}{2\pi}\int_0^{2\pi}\!d\varphi\,e^{-i\sigma m\varphi}f^\sigma({\bm k}).
\end{eqnarray}

The RS vector, calculated from Eqs.~(\ref{solf}) with $\chi$ given by (\ref{bes}), has the form
\begin{eqnarray}\label{bessel}
{\bm F}^\sigma_{ k_\perp m k_z}(\rho,\phi,z,t)=\frac{(i\sigma)^m}{\sqrt{2}k} e^{-i\sigma(\omega_k\,t- k_z
z-m\phi)}\left(\begin{array}{c} i\sigma k_-(\sigma)e^{i\sigma\phi}J_{m+1}(\xi)+
i\sigma k_+(\sigma)e^{-i\sigma\phi}J_{m-1}(\xi)
\\
k_-(\sigma)e^{i\sigma\phi}J_{m+1}(\xi)-k_+(\sigma)e^{-i\sigma\phi}J_{m-1}(\xi)
\\
k_\perp J_m(\xi)
\end{array}\right),
\end{eqnarray}
where $\xi = k_\perp\rho$ and
\begin{eqnarray}\label{qpm}
k_\pm(\sigma)=\frac{\sigma k\pm k_z}{2}.
\end{eqnarray}
In the derivation of (\ref{bessel}) we have used the following recurrence relations for the Bessel functions
\begin{subequations}\label{recur}
\begin{eqnarray}
2mJ_m(\xi)&=&\xi J_{m-1}(\xi)+\xi J_{m+1}(\xi),\\
2\partial_\xi J_m(\xi)&=&J_{m-1}(\xi)-J_{m+1}(\xi).
\end{eqnarray}
\end{subequations}
The Bessel beams (\ref{bessel}) form a complete set --- all other solutions can be obtained as linear combinations of Bessel beams. The Bessel beams are characterized by the four parameters $k_z, k_\perp, m$,
and $\sigma$. The parameters $k_z$ and $k_\perp$ ($-\infty<k_z<\infty,\; 0<k_\perp<\infty$) are the $z$ and
the $\rho$ components of the wave vector, respectively. The parameter $m$ takes on all integer values. The
components of ${\bm F}_{k_\perp m k_z }^\sigma$ in the cylindrical coordinates are simpler and allow for a
complete factoring of the angular dependence
\begin{eqnarray}\label{cylcom}
\left(\begin{array}{c}
F_\rho\\
F_\phi\\
F_z
\end{array}\right)
= \frac{(i\sigma)^m}{\sqrt{2}k} e^{-i\sigma(\omega_k\,t- k_z z - m\phi)}\left(\begin{array}{c}
i\sigma k_z\partial_\xi+ik m/\xi
\\
-\sigma k\partial_\xi - k_z m/\xi
\\
k_\perp
\end{array}\right) J_m(\xi).
\end{eqnarray}
Note, that the phase factor in front is invariant under the following simultaneous changes of the coordinates $z$ and $\phi$: $z\to z - m \alpha/k_z,\;\;\phi\to \phi + \alpha$. That means that the Bessel beam has a screw symmetry.

Bessel beams take on a very simple form in the limit, when $k_\perp\to 0$. In order to avoid the trivial (zero) result, we shall divide (for $m>0$) the expression (\ref{bes}) by $k_\perp^{m-1}$ before taking the limit
\begin{eqnarray}\label{bes0}
\lim_{k_\perp\to 0}(k_\perp^{1-m}\chi_{k_z m}^{\sigma}(\rho,\phi,z,t)
  = \frac{(i\sigma)^m}{\sqrt{2}k 2^m m!}e^{-i\sigma(\vert k_z\vert ct - k_z z)}(x+i\sigma y)^m.
\end{eqnarray}
This expression represents on the one hand an exact solution of the wave equation but on the other hand, when substituted into the Eqs.~(\ref{solf}), it also gives an approximation of the Bessel beam in the vicinity of the $z$ axis. The electromagnetic field derived from (\ref{bes0}) has some unique properties \cite{kiev}. In particular, when $m=2$, it allows for exact solutions of the equations of motion for charged particles \cite{beams}.

\section{Wave packets of Bessel beams}

Bessel beams are almost ideal representations of pencil-shaped electromagnetic radiation moving along the $z$ axis without spreading in the transverse direction. In this respect, they are a much better representation of realistic electromagnetic beams than plane waves. However, there is a price to be paid for the diffraction-free propagation: the fall-off of the field is so slow (like $1/\sqrt{\rho}$) that the energy of the electromagnetic field per unit $z$ is infinite. A standard method of avoiding the infinite energy problem for monochromatic beams consists of abandoning the requirement that the functions describing the beams are {\em exact solutions} of the Maxwell equations. Along these lines, various forms of Gaussian beams, i.e. beams with a Gaussian fall-off in the transverse direction were proposed (see \cite{oam}). We find the departure from Maxwell equations (paraxial approximation) to be too high a price to pay for saving the monochromaticity --- strictly monochromatic beams do not exist in reality, anyway. We shall show that a fast fall-off in the transverse direction can be obtained for exact solutions of Maxwell equations constructed as wave packets of Bessel beams. However, these beams will be, of course, not strictly monochromatic.

Instead of working with the RS vectors, we shall form the wave packets using the scalar functions $\chi$. Since the transformation (\ref{solf}) from $\chi$ to ${\bm F}$ is linear, the wave packets for the complete RS vector can be easily produced by taking the appropriate derivatives of $\chi$. Taking this into account, we shall often use the name ``beams'' for scalar solutions of the d'Alembert equation with the understanding that the solutions of Maxwell equations describing electromagnetic beams are to be obtained from $\chi$ according to Eqs.~(\ref{solf}).

We shall consider the most general wave packet of Bessel beams with given values of $m$ and $\sigma$. A wave packet solution of the d'Alembert equation of this type can be written as the following superposition of the functions (\ref{bes})
\begin{eqnarray}\label{genbeam}
{\chi_{m\sigma}(\rho,\phi,z,t)
 = e^{i\sigma m\phi}\int_0^\infty\!\!d k_\perp k_\perp\int_{-\infty}^\infty\!\!d k_z
g(k_z,k_\perp)e^{-i\sigma(c\sqrt{k_z^2+k_\perp^2} t - k_z z)}J_m(k_\perp\rho),}
\end{eqnarray}
where $g(k_z,k_\perp)$ is some weight function. For beam-like solutions, the longitudinal component $k_z$ should be much larger than the transverse component $k_\perp$. Note, that this condition is also essential for the paraxial approximation. In the limit, when the weight function shrinks to a product of delta functions
\begin{eqnarray}
g(k_z,k_\perp)\to \frac{(i\sigma)^m}{\sqrt{2}kk_\perp}\delta(k_z - k'_z)\delta(k_\perp - k'_\perp),
\end{eqnarray}
we obtain back the function (\ref{bes}). There are not too many weight functions that allow for an explicit evaluation of the integral (\ref{genbeam}) but there are several examples when this can be done. Not all of them will lead to well formed beams. The most important case is treated in the next section.

\section{Exact Laguerre-Gauss (LG) beams}

Exact Laguerre-Gaussian beams are most easily obtained by changing the variables in the formula (\ref{genbeam}) from $k_z$ and $k_\perp$ to $k_\pm=(\omega_k/c\pm k_z)/2=(k \pm k_z)/2$. We shall also introduce the new variables $t_\pm = t\pm z/c$ and rewrite the exponent in Eq.~(\ref{genbeam}) in the form
\begin{eqnarray}\label{newvar1}
\omega_k\,t - k_z z = c(k_+t_-+k_-t_+).
\end{eqnarray}
In these variables the general wave packet with a given values of $m$ and $\sigma$ has the form
\begin{eqnarray}\label{genbeam1}
{\chi_m^\sigma(\rho,\phi,z,t)
 = e^{i\sigma m\phi}\int_0^\infty\!\int_0^\infty\!d k_+d k_-
g(k_+,k_-)e^{-i\sigma ck_+ t_-}e^{-i\sigma ck_-t_+}J_m(2\sqrt{k_+k_-}\rho).}
\end{eqnarray}
The dependence on $t_-(t_+)$ signifies the propagation of the signal in the positive (negative) direction of the $z$ axis. In a general superposition of Bessel beams (\ref{genbeam1}), both these directions of propagation are present. However, it is clear that in a true beam-like situation one of the propagation directions must play a predominant role. We shall choose the convention that the beam propagates in the positive $z$ direction. In this case, for a true beam the dependence on $t_+$ must be weak and only cause a modulation of a propagating wave. In the limiting case, when there is no dependence on $t_+$, the wave propagates along the $z$ axis without changing its shape. In order to construct a beam-like wave packet, with $k_+$ concentrated near some large value $\Omega/c$, we shall allow only for a narrow range of values of $k_-$ as compared to $\Omega/c$.

The simplest functions $g(k_+,k_-)$ that generate exact beam-like solutions of the Maxwell equations with a Gaussian fall-off in the transverse direction have the form
\begin{eqnarray}\label{simplest}
g(k_+,k_-) = \delta(k_+-\Omega/c)k_-^{n+m/2}e^{-l^2\Omega k_-/c}\!\!,
\end{eqnarray}
where $l$, as will be shown below, determines the width of the LG beam. With this choice of $g(k_+,k_-)$, both integrations in Eq.~(\ref{genbeam1}) can be performed. Owing to a delta function, the integration over $k_+$ is immediate and it results in the replacement $k_+\to \Omega/c$. Next, we shall integrate over $k_-$ using (Eq.~6.643.4 of Ref.\cite{gr})
\begin{eqnarray}\label{grb}
\int_0^\infty\!\!d x\,x^{n+\nu/2}e^{-\alpha x}J_\nu(2\beta\sqrt{x}) = \frac{n!\beta^\nu
e^{-\beta^2/\alpha}}{\alpha^{n+\nu+1}} L_n^\nu\left(\frac{\beta^2}{\alpha}\right)\!,
\end{eqnarray}
where $L_n^m$ is the associated Laguerre polynomial. This gives
\begin{eqnarray}\label{first}
{ \int_0^\infty\!\!d k_-k_-^{n+m/2}e^{-\Omega a(t_+)k_-/c}J_m\!\!\left(\frac{2\rho\sqrt{\Omega
k_-}}{\sqrt{c}}\right) = \frac{A\rho^m}{a(t_+)^{n+m+1}}\!
\exp\left(\frac{-\rho^2}{a(t_+)}\right)L_n^m\left(\frac{\rho^2}{a(t_+)}\right),}
\end{eqnarray}
where $a(t_+)=l^2+i\sigma c^2t_+/\Omega$ and $A=n!(c/\Omega)^{n+m/2+1}$. Finally, our wave packet of Bessel beams representing an exact LG beam, expressed in Cartesian coordinates, takes on the form
\begin{eqnarray}\label{laguerre}
\chi_{\Omega nm}^\sigma(\rho,\phi,z,t) =A\frac{e^{-i\sigma\Omega(t-z/c-m\phi)}\rho^m}{a(t_+)^{n+m+1}}
\exp\left(-\frac{\rho^2}{a(t_+)}\right) L_n^m\!\left(\frac{\rho^2}{a(t_+)}\right).
\end{eqnarray}
The width of the Gaussian is equal to $l$ at the waist and grows with $t_+$ as we move away from $t_+=0$. In the optical regime, when $\Omega=10^{15}{\rm sec}^{-1}$ and $l=0.001$m, the modulation of the beam through its dependence on $t_+$ has characteristic time $l^2\Omega/c^2\approx 10^{-8}{\rm sec}^{-1}$. This time is seven orders of magnitude longer than the wave period.

In the simplest case, when $n=0$, $m=0$, the exact LG beam reduces to a pure Gaussian
\begin{eqnarray}\label{laguerre0}
\chi_{\Omega 00}^\sigma(x,y,z,t) =\frac{e^{-i\sigma\Omega(t-z/c)}}{a(t_+)}
\exp\left(-\frac{\rho^2}{a(t_+)}\right).
\end{eqnarray}
Apart from some changes in notation, this function for $\sigma=1$ coincides with the solution investigated recently by Saari, Menert, and Valtna \cite{smv} in connection with photon localization. Their solution, as it turns out, is a member of a complete set of functions, labeled by $n$ and $m$, all having an exponential fall-off in the transverse direction. There is a vast literature on this subject and various forms of the solutions of the wave equation equivalent to our formulas can be found in \cite{kis}. What makes our approach different is a systematic use of the Bessel beams as a basis from which other solutions can be obtained.

Obviously, for each function $\chi_{\Omega nm}^\sigma$ we obtain from the formulas (\ref{solf}) a solution of the full Maxwell equations. The exact LG beams fall off sufficiently fast in the transverse direction to guarantee that all physical quantities (energy, momentum, angular momentum) per unit interval in the $z$ direction are finite.

\section{Spectral decomposition of exact Laguerre-Gauss beams}

The exact Laguerre-Gauss beams given by the formula (\ref{laguerre}) are similar to the so called elegant LG beams \cite{sieg0,sieg}. However, in contrast to the elegant LG beams, the exact LG beams are not monochromatic but they solve the Maxwell equations exactly and not only in the paraxial approximation. Even though our exact LG beams are not monochromatic, they can be made nearly monochromatic by a proper choice of the parameters $\Omega$ and $l$ --- their spectrum will be sharply peaked. In order to see this, we shall decompose the function (\ref{laguerre}) into its Fourier components in the time variable
\begin{eqnarray}\label{ift}
\chi_{\omega\Omega nm}^\sigma(x,y,z) =\frac{1}{2\pi}\int_{-\infty}^\infty \!\!dt\,e^{i\omega t} \chi_{\Omega nm}^\sigma(x,y,z,t).
\end{eqnarray}
This task is made simple by returning to the original formula (\ref{genbeam1}) and observing that the integration over time leads to a delta function $\delta(c(k_++k_-)-\sigma\omega)$. The sign of $\omega$ must coincide, therefore, with the sign of $\sigma$, i.e. $\sigma={\rm sign}(\omega)$. The absolute value of $\omega$ is never less than $\Omega$ because $k_+=\Omega/c$ and $k_-\geq 0$. The final expression for the Fourier transform, obtained after using the two delta functions, can be written as a product of the spectral weight function $w(\omega)$ and the time-independent part of the Bessel beam
\begin{eqnarray}\label{prod}
\chi_{\omega\Omega nm}(\rho,\phi,z) = w(\omega)e^{i k_z z/c}e^{i m\phi} J_m(k_\perp\rho),
\end{eqnarray}
where the spectral weight function is
\begin{eqnarray}\label{weight}
w(\omega)=C\theta(\vert\omega\vert-\Omega)\, k_-^{n+m/2}e^{-l^2\Omega k_-/c},
\end{eqnarray}
$C$ is a constant, and the parameters $k_z,k_\perp$ and $k_-$ are now the following functions of $\Omega$ and $\omega$
\begin{eqnarray}\label{par}
k_z=(2\Omega-\vert\omega\vert)/c,\;\; k_\perp=2\sqrt{(\vert\omega\vert-\Omega)\Omega}/c,\;\;
k_-=(\vert\omega\vert-\Omega)/c.
\end{eqnarray}
In order to make the comparison of different cases easier, we will normalize all weight functions to 1
\begin{eqnarray}\label{norm1}
\int d\omega\,w(\omega) = 1,
\end{eqnarray}
which changes the value of the constant $C$ to
\begin{eqnarray}\label{norm2}
C = \frac{\left(c^2/l^2\Omega\right)^{n+m/2+1}}{\Gamma(n+m/2+1)}.
\end{eqnarray}
\begin{figure}
\centering
\includegraphics[width=0.48\textwidth]{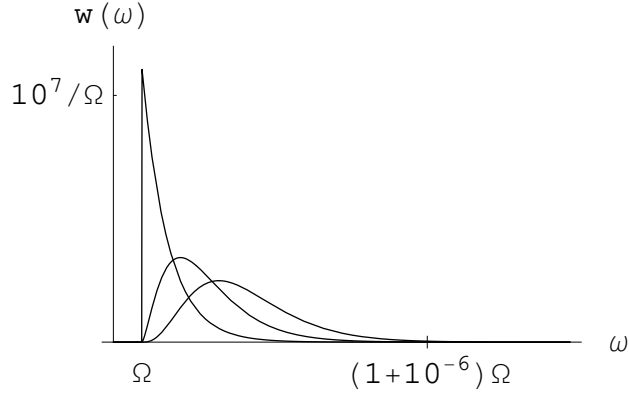}
\caption{The normalized spectral weight $w(\omega)$ plotted as a function of $\omega$ for the following values of the parameters: $\sigma=1$, $\Omega = 10^{15}$s$^{-1}$ and $l=0.001$m. The sharpest peak represents the case $n=0,m=0$ and the remaining two correspond to $n=1,m=1$ and $n=2,m=2$. As we increase these numbers, the peaks flatten more and more but they are always localized near $\Omega$.}\label{fig}
\end{figure}
The spectral weight function of an exact LG beam has a cutoff at $\vert\omega\vert=\Omega$ and is concentrated near this frequency. The prefactor $k_-^{n+m/2}$ shifts the maximum from $\Omega$ to $\Omega+c(n+m/2)/2l$. In Fig.~\ref{fig} we show the normalized weight functions for three sets of numbers $n$ and $m$.

\section{Quantized electromagnetic field}\label{qef}

In quantum electrodynamics the RS vector and its complex conjugate are replaced by the field operators built from the operators of the electric and magnetic field vectors
\begin{eqnarray}
{\hat{\bm F}}({\bm r},t)=\sqrt{\frac{\epsilon_0}{2}}\left({\hat{\bm E}}({\bm r},t)+i c{\hat{\bm B}}({\bm
r},t)\right).
\end{eqnarray}
The canonical quantization of the electromagnetic field that employs the mode expansion into a set of harmonic oscillators \cite{louisell,enc} yields the following equal-time commutation relation between these field operators
\begin{eqnarray}\label{ccr0}
\left[{\hat F}_i({\bm r},t),{\hat F}^\dagger_j({\bm r}',t)\right] = -\hbar c\epsilon_{ijk}\partial_k\delta^{(3)}({\bm r}-{\bm r}'),\; \left[{\hat F}_i({\bm r},t),{\hat F}_j({\bm r}',t)\right] = 0,\;\left[{\hat F}^\dagger_i({\bm r},t),{\hat F}^\dagger_j({\bm r}',t)\right] = 0.\;
\end{eqnarray}

The expansion of the RS operator into plane waves is obtained from Eq.~(\ref{super1}) by replacing the classical amplitudes $f^\pm({\bm k})$ by the creation and annihilation operators $f^+({\bm k})\to\sqrt{\hbar}\,{\hat a}({\bm k})$ and $f^-({\bm k})\to \sqrt{\hbar}\,{\hat b}^\dagger({\bm k})$
\begin{eqnarray}\label{superq}
{\hat{\bm F}}({\bm r},t)=\sqrt{\hbar}\int\!\!d{\bm k}\,{\bm e}({\bm k})\left({\hat a}({\bm k})e^{-i\omega
t+i{\bm k}\cdot{\bm r}}\!+{\hat b}^\dagger({\bm k})e^{i\omega t-i{\bm k}\cdot{\bm r}}\right).
\end{eqnarray}
The amplitude  $f^-({\bm k})$ is being replaced by the creation operator because it is multiplied by the opposite frequency factor $\exp(i\omega t)$. Commutation relations between the creation and annihilation operators
\begin{eqnarray}\label{ccr1}
\left[{\hat a}({\bm k}),{\hat a}^\dagger({\bm k}')\right]=(2\pi)^3\omega\,\delta^{(3)}({\bm k}-{\bm k}'),\;\;\left[{\hat b}({\bm k}),{\hat b}^\dagger({\bm k}')\right]=(2\pi)^3\omega\,\delta^{(3)}({\bm k}-{\bm k}'),
\end{eqnarray}
with all remaining commutators vanishing, reproduce correctly Eqs.~(\ref{ccr0}) since
\begin{eqnarray}\label{idn}
e_i({\bm k})e^*_j({\bm k})-e^*_i({\bm k})e_j({\bm k}) = i\epsilon_{ijk}n_k.
\end{eqnarray}
The creation and annihilation operators in this work differ by a factor $\sqrt{(2\pi)^3\omega}$ from the traditional ones that satisfy the commutation relation (\ref{ccr1}) with just a straight delta function on the right hand side. We have departed from the standard normalization of these operators for two reasons. First, we wanted to keep a close connection between the classical amplitudes and the annihilation/creation operators. Second, as shown in \cite{qed}, our normalization makes the photon density operator in momentum space ${\hat a}^\dagger({\bm k}){\hat a}({\bm k})$ a relativistic scalar.

The commutation relations (\ref{ccr1}) enforce the replacement of the classical amplitude $f^-({\bm k})$, that multiplies the negative frequency solution, by the creation operator ${\hat b}^\dagger({\bm k})$. Replacing it by the annihilation operator would yield the wrong sign in the commutation relations (\ref{ccr1}). The operators ${\hat a}^\dagger({\bm k})$ and ${\hat b}^\dagger({\bm k})$ create different photons (of opposite circular polarization). The fact that the RS field operator contains only ${\hat a}({\bm k})$ and ${\hat b}^\dagger({\bm k})$ (and not ${\hat b}({\bm k})$ and ${\hat a}^\dagger({\bm k})$) leads to some unique properties of this field operator which are different from those of the electric and magnetic field operators. To see this, let us consider a general state of a single photon, created from the vacuum by an arbitrary linear combination of creation operators ${\hat{a}}^\dagger$ and ${\hat{b}}^\dagger$
\begin{eqnarray}\label{arb}
{\hat a}^\dagger_\psi = \int\!\frac{d{\bm k}}{\omega}\,\left(\psi^+({\bm k}){\hat{a}}^\dagger({\bm
k})+\psi^-({\bm k}){\hat{b}}^\dagger({\bm k})\right).
\end{eqnarray}
We have to impose the normalization condition
\begin{eqnarray}\label{norm}
\int\!\frac{d{\bm k}}{\omega}\,\left(\vert\psi^+({\bm k})\vert^2+\vert\psi^-({\bm k})\vert^2\right)=1
\end{eqnarray}
to guarantee proper commutation relations between ${\hat a}_\psi$ and ${\hat a}^\dagger_\psi$. With the use of this creation operator we may define a coherent state of the electromagnetic field
\begin{eqnarray}\label{coh}
\vert\psi\rangle_{\rm c} = \exp\left({-\frac{1}{2}\langle{\hat N}\rangle}\right)\exp\left({\sqrt{\langle{\hat
N}\rangle}\,{\hat a}_\psi^\dagger}\right)\vert 0\rangle,
\end{eqnarray}
where $\langle{\hat N}\rangle$ is the average number of photons in the coherent state. The expectation value of the field operator in a coherent state is given by the formula
\begin{eqnarray}\label{avercoh}
\langle{\hat{\bm F}}({\bm r},t)\rangle_{\rm c}=\sqrt{\hbar\langle{\hat N}}\rangle\int\!d^3\!k\;{\bm e}({\bm k})\left(\psi^+({\bm k})e^{-i\omega t+i{\bm k}\cdot{\bm r}}\!+(\psi^{-}({\bm k}))^*e^{i\omega t-i{\bm k}\cdot{\bm r}}\right).
\end{eqnarray}
Note the complex conjugation of the wave function in the second term. This is caused by the fact that this term has opposite sign of the frequency (negative energy in the quantum-mechanical setting). Every mode function in the expansion of the RS vector with negative energy must be interpreted as the complex conjugate of the photon wave function.

Since coherent states are the right eigenstates of the annihilation operators and the left eigenstates of the creation operators
\begin{eqnarray}\label{lr}
{\hat{a}}({\bm k})\vert\psi\rangle_{\rm c}=\psi^+({\bm k})\vert\psi\rangle_{\rm c},\;\;\; _{\rm
c}\langle\psi\vert{\hat{b}}^\dagger({\bm k})={}_{\rm c}\langle\psi\vert\,\psi^{-*}({\bm k}),
\end{eqnarray}
the expectation value of {\em any product} of the RS field operators is equal to the product of the corresponding classical RS vectors. One can show with the use of (\ref{lr}) that the expectation value of any departure from the average value vanishes
\begin{eqnarray}\label{cohav}
\langle({\hat{\bm F}}({\bm r},t)-\langle{\hat{\bm F}}({\bm r},t)\rangle_{\rm c})^n\rangle_{\rm c}=0.
\end{eqnarray}
Note, that this property holds just for the RS operator (and also for its hermitian conjugate). It is so, because the RS operator contains the annihilation operators of right-handed photons and the creation operators of left-handed photons. The electric and magnetic field vectors exhibit always (even for coherent states) large fluctuations around the average value.

The formula (\ref{avercoh}) gives also an inverse connection between a classical RS vector and a quantum coherent state. From a solution of classical Maxwell equations, we may find the coefficients in an expansion of the RS vector into plane waves. Their norm determines the average photon number $\langle{\hat N}\rangle$
while the normalized coefficients $\psi_\pm({\bm k})$ determine the creation operator (\ref{arb}). Together, they enable us to define the coherent state (\ref{coh}). Obviously, the average value of the RS field operator in this coherent state reproduces the original solution of the Maxwell equations.

Since in this paper we are interested in the properties of beams carrying angular momentum, we shall now use the expansion of the operator ${\hat{\bm F}}({\bm r},t)$ into the Bessel beams. Following the same steps as in the classical case, we obtain
\begin{eqnarray}\label{bessq1}
{\hat{\bm F}}({\bm r},t) = \sqrt{\hbar}\sum_{m=-\infty}^{\infty}\int_{-\infty}^{\infty}\!\!\frac{d
k_z}{2\pi}\int_{0}^{\infty}\!\frac{d k_\perp k_\perp}{2\pi} \Big({\bm F}_{k_\perp m
k_z}^{+}(\rho,\phi,z,t){\hat a}(k_\perp,m,k_z) +{\bm F}_{k_\perp m k_z}^{-}(\rho,\phi,z,t){\hat
b}^\dagger(k_\perp,m,k_z)\Big),
\end{eqnarray}
where the new creation and annihilation operators are defined as
\begin{subequations}
\begin{eqnarray}\label{annihm}
{\hat a}^\dagger(k_\perp,m,k_z) = \frac{1}{2\pi}\int_0^{2\pi}d\varphi\,e^{i m\varphi}{\hat a}^\dagger({\bm k}),\;\; {\hat a}(k_\perp,m,k_z) = \frac{1}{2\pi}\int_0^{2\pi}d\varphi\,e^{-i m\varphi}{\hat a}({\bm k}),\\
{\hat b}^\dagger(k_\perp,m,k_z) = \frac{1}{2\pi}\int_0^{2\pi}d\varphi\,e^{i m\varphi}{\hat b}^\dagger({\bm k}),\;\; {\hat b}(k_\perp,m,k_z,) = \frac{1}{2\pi}\int_0^{2\pi}d\varphi\,e^{-i m\varphi}{\hat b}({\bm k}).
\end{eqnarray}
\end{subequations}
Thus, ${\bm F}_{k_\perp m k_z}^{\pm}$ play the role of mode functions. According to the standard interpretation of field operators in relativistic quantum field theory, ${\bm F}_{k_\perp m k_z}^{+}$ is the wave function ${\bm\psi}_{k_\perp m k_z}^{+}$ of the photon annihilated by ${\hat a}(k_\perp,m,k_z)$ and ${\bm F}_{k_\perp m k_z}^{-}$ is the complex conjugate of the wave function ${\bm\psi}_{k_\perp m k_z}^{-}$ of the photon created by ${\hat b}^\dagger(k_\perp,m,k_z)$.
\begin{eqnarray}\label{wf}
{\bm F}_{k_\perp m k_z}^{+}(\rho,\phi,z,t)={\bm\psi}_{k_\perp m k_z}^{+}(\rho,\phi,z,t)=\frac{i^m}{\sqrt{2}k} e^{-i(\omega_k\,t- k_z z- m\phi)}\left(\begin{array}{c} i k_-e^{i\phi}J_{m+1}(\xi)+i k_+e^{-i\phi}J_{m-1}(\xi)\\
k_-e^{i\phi}J_{m+1}(\xi)-k_+e^{-i\phi}J_{m-1}(\xi)\\
k_\perp J_m(\xi)
\end{array}\right),\\
\left({\bm F}_{k_\perp m k_z}^{-}(\rho,\phi,z,t)\right)^*={\bm\psi}_{k_\perp m k_z}^{-}(\rho,\phi,z,t)=\frac{i^m}{\sqrt{2}k} e^{-i(\omega_k\,t- k_z z- m\phi)}\left(\begin{array}{c} -i k_+e^{i\phi}J_{m+1}(\xi)-i k_-e^{-i\phi}J_{m-1}(\xi)\\
-k_+e^{i\phi}J_{m+1}(\xi)+k_-e^{-i\phi}J_{m-1}(\xi)\\
k_\perp J_m(\xi)
\end{array}\right).
\end{eqnarray}
The notion of the photon wave function has been discussed in detail in our review paper \cite{pio} and also, more recently, in the report by Keller \cite{ok}. In the next section we shall analyze the quantum mechanical properties of the single-photon states described by the wave functions ${\bm\psi}_{k_\perp m k_z}^{\pm}$.

\section{Properties of photon wave functions}

We have constructed in the previous section photon wave functions using the apparatus of quantum field theory. However, there is a more pedestrian way that may be useful to better explain this construction. Let us assume that we want to repeat the success of Schr\"odinger and we want to construct a wave mechanics of photons. We shall need photon wave functions ${\bm\psi}$ satisfying an analog of the Schr\"odinger equation and a set of operators representing basic physical quantities. Since photons have spin one, we need a vector wave function. The general form of the Schr\"odinger equation is
\begin{eqnarray}\label{sch}
i\hbar\partial_t{\bm\psi}({\bm r},t)={\hat H}{\bm\psi}({\bm r},t)
\end{eqnarray}
and all we need is a photon Hamiltonian. The operators representing physical quantities associated with the generators of basic transformations are easily constructed. The momentum operator (associated with an infinitesimal displacement) and the angular momentum (associated with an infinitesimal rotation) have the form
\begin{eqnarray}\label{gen}
{\hat{\bm p}} = \frac{\hbar}{i}{\bm\nabla},\;\;\; {\hat{\bm M}} = \frac{\hbar}{i}{\bm r}\times{\bm\nabla} +
\hbar{\hat{\bm s}},
\end{eqnarray}
where ${\hat{\bm s}}$ stands for a vector built from the spin-one matrices,
\begin{eqnarray}\label{spin1}
{\hat s}_x=\left(\begin{array}{ccc}
0 & 0 & 0\\
0 & 0 & -i\\
0 & i & 0\end{array}\right),\;\; {\hat s}_y=\left(\begin{array}{ccc}
0 & 0 & i\\
0 & 0 & 0\\
-i & 0 & 0\end{array} \right),\;\; {\hat s}_z=\left(\begin{array}{ccc}
0 & -i & 0\\
i & 0 & 0\\
0 & 0 & 0\end{array} \right).
\end{eqnarray}
We shall also need the helicity operator ${\hat\Lambda}$  --- the sign of the projection of the angular momentum on the momentum.
\begin{eqnarray}\label{hel}
{\hat\Lambda}=\frac{1}{\sqrt{{\hat p}_x^2+{\hat p}_y^2+{\hat p}_z^2}}({\hat{\bm s}}\!\cdot\!{\hat{\bm p}}).
\end{eqnarray}

It has been noticed seventy five years ago (cf. Ref.\cite{pio} for a historical review) and rediscovered over and over again, that the complex form of Maxwell equations (\ref{max}) with the help of the relation
\begin{eqnarray}\label{rel}
{\bm\nabla}\times=-i({\hat{\bm s}}\cdot{\bm\nabla})
\end{eqnarray}
can be cast into a Schr\"odinger-like form
\begin{eqnarray}\label{schmaj}
i\hbar\partial_t{\bm\psi}({\bm r},t)=c({\hat{\bm s}}\!\cdot\!{\hat{\bm p}}){\bm\psi}({\bm r},t),
\end{eqnarray}
with the rescaled helicity operator playing the role of a Hamiltonian ${\hat H}=c({\hat{\bm
s}}\!\cdot\!{\hat{\bm p}})$. Thus, there is a close relation between the helicity and the Hamiltonian. However, one must exercise care because the Maxwell equations have solutions with positive and negative frequencies, while photons always have positive energy. Therefore, solutions of the complex Maxwell equations with negative frequencies must be interpreted as complex conjugate wave functions.

We shall now analyze the wave functions ${\bm\psi}_{k_\perp m k_z}^{\pm}$ associated with the Bessel beams using the tools provided by wave mechanics. To this end we choose the following set of four commuting operators: the component of the momentum in the $z$ direction ${\hat p}_z$, the square of the transverse momentum ${\hat p}_x^2+{\hat p}_y^2$, the component of the angular momentum in the $z$ direction ${\hat M}_z$, and the helicity ${\hat\Lambda}$. It is only a matter of straightforward calculations to check that both wave functions ${\bm\psi}_{k_\perp m k_z}^{\pm}$ are eigenfunctions of the first three operators belonging to the eigenvalues $\hbar k_z$, $\hbar^2k_\perp^2$, and $\hbar m$, respectively. To verify that they are also eigenfunctions of the helicity operator, no additional calculations are needed. Both functions ${\bm\psi}_{k_\perp m k_z}^{\pm}$ satisfy the complex form of Maxwell equations, but they have opposite signs of the frequency $\omega_k$. Therefore, we obtain ${\bm\nabla}\times{\bm\psi}^{\pm}=\pm k{\bm\psi}^{\pm}$, and since there is no imaginary unit in these equations, the same relation holds for complex conjugate functions. With the use of (\ref{rel}), we obtain
\begin{eqnarray}\label{fin}
 {\hat\Lambda}{\bm\psi}_{k_\perp m k_z}^{\pm}=\frac{\hat H}{\vert\hat H\vert}{\bm\psi}_{k_\perp m k_z}^{\pm}=\frac{({\hat{\bm s}}\!\cdot\!{\hat{\bm p}})}{\hbar k}{\bm\psi}_{k_\perp m k_z}^{\pm}=\pm{\bm\psi}_{k_\perp m k_z}^{\pm}.
\end{eqnarray}
Thus, according to the discussion in the previous section, a classical Bessel beam (\ref{bessel}) is made of photons with the quantum numbers $\hbar k_z$, $\hbar^2k_\perp^2$, $\hbar m$, and $\sigma$.

There is one problem with the photon wave function that for many physicists invalidates this notion altogether. Namely, the standard Born interpretation of the photon wave function ${\bm\psi}({\bm r},t)$ fails because there is no well defined position operator. This problem, however, is common to all particles described by relativistic quantum mechanics and it simply follows from the fact that a positive energy solution ${\bm\psi}^+({\bm r},t)$ of any relativistic wave equation loses this property when it is multiplied by ${\bm r}$. In particular, we encounter this problem for electrons described by the Dirac equation. In this case, however, the probabilistic interpretation is based on the charge distribution. Since photons do not carry any charge, we have to base the probabilistic interpretation on the energy distribution.

The RS vectors can serve as photon wave functions to calculate an expectation values of a quantum-mechanical operator ${\hat O}$ provided this expectation value is defined in the following way \cite{good,pio}
\begin{eqnarray}\label{exp}
\langle{\hat O}\rangle=N^{-1}\int\!d^3r{\bm F}^*({\bm r},t)\cdot\frac{1}{\sqrt{\vert\hat H\vert}}{\hat
O}\frac{1}{\sqrt{\vert\hat H\vert}}{\bm F}({\bm r},t).
\end{eqnarray}
where the absolute value takes care of the fact that the Hamiltonian has both positive and negative eigenvalues and the norm $N$ is defined as
\begin{eqnarray}\label{norm3}
N=\int\!d^3r{\bm F}^*({\bm r},t)\cdot\frac{1}{\vert\hat H\vert}{\bm F}({\bm r},t).
\end{eqnarray}
This prescription can be intuitively understood as follows: The energy density ${\bm F}^*\!\cdot\!{\bm F}$ must be ``divided'' by the energy ${\vert\hat H\vert}$ to produce a ``probability density''.

The division by the energy can be implemented in Fourier space as a division by $\hbar\omega_k$. The resulting expression, when transformed to coordinate space has the form \begin{eqnarray}\label{photnum}
N=\frac{1}{2\pi^2\hbar\,c}\int\!d^3r\int\!d^3r'{\bm F}^*({\bm r},t)\!\cdot\!\frac{1}{\vert{\bm r}-{\bm
r}'\vert^2}{\bm F}({\bm r}',t).
\end{eqnarray}
Using the formula (\ref{exp}) one can check that the correspondence principle is satisfied for the energy, momentum, and angular momentum, namely
\begin{subequations}
\begin{eqnarray}
\langle{\hat H}\rangle&=&\frac{1}{N}\int\!d^3r\,{\bm F}^*\!\cdot\!{\bm F},\label{exp1}\\
\langle{\hat{\bm p}}\rangle&=&\langle-i\hbar{\bm\nabla}\rangle=\frac{-i}{c N}\int\!d^3r\,{\bm F}^*\times{\bm F},\label{exp2}\\
\langle{\hat{\bm M}}\rangle&=&\langle-i\hbar\,{\bm r}\times{\bm\nabla} + \hbar\,{\hat{\bm
s}}\rangle=\frac{-i}{c N}\int\!d^3r\,\left({\bm r}\times({\bm F}^*\times{\bm F})\right).\label{exp3}
\end{eqnarray}
\end{subequations}
The formula for the expectation value of the helicity, like the one for the norm, has a nonlocal character
\begin{eqnarray}\label{avghel}
\langle{\hat{\bm\Lambda}}\rangle&=&\left\langle\frac{\hat H}{\vert\hat
H\vert}\right\rangle=\frac{1}{2\pi^2\hbar c N}\int\!d^3r\int\!d^3r'{\bm F}^*({\bm
r},t)\!\cdot\!\frac{1}{\vert{\bm r}-{\bm r}'\vert}{\bm\nabla}\times{\bm F}({\bm r}',t).
\end{eqnarray}
This double integral has been obtained by Deser and Teiteltboim \cite{dt} as a generator of dual rotations in classical electrodynamics.

Let us suppose now that ${\bm F}$ is an arbitrary eigenfunction of ${\hat M}_z=-i\hbar(x\partial_y-y\partial_x)+\hbar{\hat s_z} $ belonging to an eigenvalue $\hbar m$. In this case, from Eq.~(\ref{exp3}) we obtain
\begin{eqnarray}
\langle{\hat{M}_z}\rangle=\frac{-i}{c}\int\!d^3r\,\left({\bm r}\times({\bm F}^*\times{\bm F})\right)_z=N\hbar\,m\label{res}.
\end{eqnarray}
The number $N$, playing the role of the norm in quantum mechanics of the photon, acquires a physical interpretation in quantum electrodynamics \cite{zeld,ibb2,ibb3}. Namely, it is the average number of photons $N=\langle{\hat N}\rangle$ in a coherent state associated with the RS vector ${\bm F}$. Hence, the value of the angular momentum obtained from the classical formula is equal to the quantum eigenvalue $\hbar\,m$ (i.e. the value per one photon) multiplied by the number of photons.

\section{Beams in momentum representation}\label{momr}

Momentum representation, well known from nonrelativistic quantum mechanics, is very useful in the description of photons since it is free of the difficulties with the probabilistic interpretation encountered in the coordinate representation. Actually, we have already introduced the momentum representation by writing down the formula (\ref{super1}) for the RS vector. The amplitudes $\psi^+({\bm k})=f^+({\bm k})$ and $\psi^-({\bm k})=(f^-({\bm k}))^*$ may be viewed as the photon wave functions in momentum representation. Complex conjugation in the second formula, as we have explained in Section \ref{qef}, is due to the fact that the amplitude $f^-({\bm k})$ is multiplied by the plane wave factor with opposite frequency. The amplitudes $\psi^\pm$ carry full information about the RS vector. Every operator ${\hat O}$ acting on the left hand side of Eq.~(\ref{super1}) can be translated into an operator in the momentum representation acting on the amplitudes $\psi^\pm({\bm k})$. In particular, an infinitesimal displacement ${\bm\nabla}$ acting on ${\bm F}({\bm r},t)$ is reproduced by a multiplication of $\psi^\pm({\bm k})$ by $i{\bm k}$. Therefore, the momentum operator ${\hat{\bm p}}$ acts on $\psi^\pm({\bm k})$ simply as a multiplication by $\hbar{\bm k}$, as is always the case in momentum representation. The action of an infinitesimal rotation ${\bm r}\times{\bm\nabla}+i{\hat{\bm s}}$ on ${\bm F}({\bm r},t)$ translated into an action on $\psi^\pm({\bm k})$ is not unique. Its form depends on the choice of the polarization vector. For our choice of ${\bm e}({\bm k})$, that resulted from the Whittaker representation (\ref{solf}), we obtain the following expressions for the components of the angular momentum operator
\begin{subequations}\label{angmomf}
\begin{eqnarray}
 {\hat M}_x \psi^\pm({\bm k})&=&\hbar\left(-i({\bm k}\times{\bm\partial}_{\bm k})_x\pm\frac{k k_x}{k_x^2+k_y^2}\right)\psi^\pm({\bm k}),\label{angmomx}\\
 {\hat M}_y \psi^\pm({\bm k})&=&\hbar\left(-i({\bm k}\times{\bm\partial}_{\bm k})_y\pm \frac{k k_y}{k_x^2+k_y^2}\right)\psi^\pm({\bm k}),\label{angmomy}\\
 {\hat M}_z \psi^\pm({\bm k})&=&-i\hbar({\bm k}\times{\bm\partial}_{\bm k})_z \psi^\pm({\bm k}).\label{angmomz}
\end{eqnarray}
\end{subequations}
The Whittaker representation is very convenient because in this representation ${\hat M}_z$ reduces just to the $z$ component of the ordinary orbital angular momentum so that the eigenstates of ${\hat M}_z$ depend on the polar angle $\varphi$ through the phase factor $\exp(im\varphi)$. This may look counterintuitive because one would expect also a contribution from the photon spin. The absence of such a contribution is a consequence of the special choice of the phase of ${\bm e}({\bm k})$ and hence of the phase of $f^\pm({\bm k})$. Another natural choice of the complex Hertz potential ${\bm Z}$, mentioned in Section \ref{two}, leads to the formulas
\begin{subequations}\label{angmomf1}
\begin{eqnarray}
{\hat M}_x \psi^\pm({\bm k})&=&\hbar\left(-i({\bm k}\times{\bm\partial}_{\bm k})_x\pm\frac{k_x}{k+k_z}\right)\psi^\pm({\bm k}),\label{angmomx1}\\
{\hat M}_y \psi^\pm({\bm k})&=&\hbar\left(-i({\bm k}\times{\bm\partial}_{\bm k})_y\pm \frac{k_y}{k+k_z}\right)\psi^\pm({\bm k}),\label{angmomy1}\\
{\hat M}_z \psi^\pm({\bm k})&=&\hbar\left(-i({\bm k}\times{\bm\partial}_{\bm k})_z \pm 1\right)\psi^\pm({\bm
k}).\label{angmomz1}
\end{eqnarray}
\end{subequations}
Here, the form of $M_z$ agrees with our expectations. There is a contribution from the photon spin equal to $\pm\hbar$, but it was manufactured by choosing a particular phase of ${\bm e}({\bm k})$.

In summary, only the total angular momentum of the photon has a physical meaning (as a generator of rotations) while the separation into the orbital and the spin part is not unique. The projection of the spin part on the direction of momentum --- helicity --- (orbital angular momentum does not contribute here) is also well defined. For all choices of ${\bm e}({\bm k})$, the action of the helicity operator in momentum representation, namely  ${\hat\Lambda}={\bm k}\!\cdot\!{\hat{\bm M}}/\hbar k$, results in ${\hat\Lambda}\psi^\pm=\pm \psi^\pm$.

Now, we apply these results to the description of the Bessel beams and exact LG beams. In order to make a distinction between the labels and the arguments, we introduce a primed vector ${\bm k}'$ to denote the arguments (integration variables in the plane-wave representation) of the functions $f^\pm$. We retain the vector ${\bm k}$ for labeling the quantum numbers of a beam. For the Bessel beams we obtain the amplitudes $f^\pm({\bm k}')$ in the form
\begin{subequations}\label{fbessel}
\begin{eqnarray}
f^+_{k_\perp m k_z}(k_\perp',\varphi',k_z')
&=& (-i)^m \sqrt{2}(2\pi)^2 k'e^{i m\varphi'}\delta(k_\perp'-k_\perp)\delta(k_z'-k_z),\label{fbesselp}\\
f^-_{k_\perp m k_z}(k_\perp',\varphi',k_z')
 &=& (f^+_{k_\perp m k_z}(k_\perp',\varphi',k_z'))^*.\label{fbesselm}
\end{eqnarray}
\end{subequations}
Indeed, upon the substitution of (\ref{fbesselp}) into the formula (\ref{super1}) and after evaluating the integrals with respect to $k_z$ and $k_\perp$, we can use the standard integral representation of the Bessel function to obtain
\begin{eqnarray}\label{fbessel1}
\int_0^{2\pi}\!\!\!d\varphi\, e^{i(m\varphi+k_\perp\rho\cos(\varphi-\phi))}=2\pi i^me^{i m\phi}
J_m({k_\perp\rho}).
\end{eqnarray}
The formula for $\sigma=-1$ is obtained by complex conjugation. The amplitudes $f^\pm$ for the exact LG beam can be obtained by convoluting the expressions (\ref{fbessel}) with (\ref{simplest}) and they have the form
\begin{subequations}\label{flg}
\begin{eqnarray}
f^+_{\Omega nm}(k_+',\varphi',k_-') &=& (-i)^m \sqrt{2} (2\pi)^2 k' e^{i m\varphi'}\delta(k_+'-\Omega/c)
k_-'^{n+m/2}e^{-l^2\Omega k_-'/c},\label{wavefp}\\
f^-_{\Omega nm}(k_+',\varphi',k_-') &=& (f^+_{\Omega nm}(k_+',\varphi',k_-'))^*.\label{wavefm}
\end{eqnarray}
\end{subequations}
Since the mode functions for the Bessel beams (\ref{fbessel}) and Laguerre-Gauss beams (\ref{flg}) describing opposite polarizations are related by complex conjugation, the corresponding wave functions are identical, as was to be expected.

\section{Conclusions}

We have built exact solutions of Maxwell equations carrying angular momentum as superpositions of Bessel beams. In particular, we analyzed the spectral properties of the exact Laguerre-Gauss beams. We have shown that we may associate photon wave functions in the form of the Riemann-Silberstein vector with all classical solutions of Maxwell equations. The Whittaker representation of the RS vector, that enables one to condense full solutions of the Maxwell equations into just one complex function, turned out to be particularly useful in this context. The {\it quantum numbers} of the photon states may serve as {\it labels} characterizing the classical solutions. The photon states with a prescribed angular momentum, such as the Bessel and Laguerre-Gauss beams, replace the commonly used plane waves and are best suited to describe light beams. The photon wave functions also appear in quantum electrodynamics as the mode functions in the expansion of the electromagnetic field operator into creation and annihilation operator. We have demonstrated that the RS field operator has a unique property: in coherent states the expectation value of the $n$-th power of the RS operator is equal to the $n$-th power of the expectation value of this operator. This means that, in contrast to the electric and magnetic fields, the RS field operator exhibits no fluctuations around the coherent state average value. We identified the source of the difficulties in the splitting of the total angular momentum into the orbital and the spin parts. The nonuniqueness of this splitting is caused by an arbitrariness in the choice of the phase factor of the photon wave function in momentum representation. This, in turn, may be linked to the gauge freedom of the complex Hertz vector.

\section*{Acknowledgments}

This research has been partly supported by the Polish Ministry of Scientific Research Grant Quantum Information and Quantum Engineering.

\end{document}